\documentclass{emulateapj}
\usepackage{amsmath,amssymb}
\usepackage{graphicx}
\usepackage{aas_macros}

\usepackage[normalem]{ulem}
\usepackage{color}

\usepackage{subfigure}

\slugcomment{Accepted for publication in the Astrophysical Journal Letters}

\shorttitle{$^{13}$CO/C$^{18}$O Across Galaxy Disks}
\shortauthors{Jim\'enez-Donaire et al.}

\begin{document}

\title{$^{13}$CO/C$^{18}$O Gradients Across the Disks of Nearby Spiral Galaxies}
    \author{Mar\'ia J. Jim\'enez-Donaire\altaffilmark{1}, Diane Cormier\altaffilmark{1}, Frank Bigiel\altaffilmark{1}, Adam K. Leroy\altaffilmark{2}, Molly Gallagher\altaffilmark{2}, Mark. R. Krumholz\altaffilmark{3}, Antonio Usero\altaffilmark{4}, Annie Hughes\altaffilmark{5,6}, Carsten Kramer\altaffilmark{7}, David Meier\altaffilmark{8}, Eric Murphy\altaffilmark{9}, J\'er\^ome Pety\altaffilmark{10,11}, Eva Schinnerer\altaffilmark{12}, Andreas Schruba\altaffilmark{13}, Karl Schuster\altaffilmark{10}, Kazimierz Sliwa\altaffilmark{12}, Neven Tomicic\altaffilmark{12}}

\altaffiltext{1}{Institut f\"ur theoretische Astrophysik, Zentrum f\"ur Astronomie der Universit\"at Heidelberg, Albert-Ueberle Str. 2, 69120 Heidelberg, Germany; m.jimenez@zah.uni-heidelberg.de}
\altaffiltext{2}{Department of Astronomy, The Ohio State University, 140 W 18$^{\rm th}$ St, Columbus, OH 43210, USA}
\altaffiltext{3}{Research School of Astronomy \& Astrophysics, Australian National University, Canberra, ACT 2611, Australia}
\altaffiltext{4}{Observatorio Astron\'omico Nacional, Alfonso XII 3, 28014, Madrid, Spain}
\altaffiltext{5}{CNRS, IRAP, 9 Av. colonel Roche, BP 44346, F-31028 Toulouse cedex 4, France}
\altaffiltext{6}{Universit\'{e} de Toulouse, UPS-OMP, IRAP, F-31028 Toulouse cedex 4, France}
\altaffiltext{7}{Instituto de Astrof\'isica de Andaluc\'ia IAA-CSIC, Glorieta de la Astronom\'ia s/n, E-18008, Granada, Spain}
\altaffiltext{8}{Department of Physics, New Mexico Institute of Mining and Technology, 801 Leroy Place, Soccoro, NM 87801, USA}
\altaffiltext{9}{National Radio Astronomy Observatory, 520 Edgemont Road, Charlottesville, VA 22903, USA}
\altaffiltext{10}{Institut de Radioastronomie Millim\'etrique (IRAM), 300 Rue de la Piscine, F-38406 Saint Martin d'H\`eres, France}
\altaffiltext{11}{Observatoire de Paris, 61 Avenue de l'Observatoire, F-75014 Paris, France}
\altaffiltext{12}{Max-Planck-Institut f\"ur Astronomie, K\"onigstuhl 17, 69117 Heidelberg, Germany}
\altaffiltext{13}{Max-Planck-Institut f\"ur extraterrestrische Physik, Giessenbachstrasse 1, 85748 Garching, Germany}

\begin{abstract}

We use the IRAM Large Program EMPIRE and new high-resolution ALMA data to measure $^{13}$CO(1-0)/C$^{18}$O(1-0) intensity ratios across nine nearby spiral galaxies. These isotopologues of $^{12}$CO are typically optically thin across most of the area in galaxy disks, and this ratio allows us to gauge their relative abundance due to chemistry or stellar nucleosynthesis effects. Resolved $^{13}$CO/C$^{18}$O gradients across normal galaxies have been rare due to the faintness of these lines. We find a mean $^{13}$CO/C$^{18}$O ratio of 6.0$\pm$0.9 for the central regions of our galaxies. This agrees well with results in the Milky Way, but differs from results for starburst galaxies (3.4$\pm$0.9) and ultraluminous infrared galaxies (1.1$\pm$0.4). In our sample, the $^{13}$CO/C$^{18}$O ratio consistently increases with increasing galactocentric radius and decreases with increasing star formation rate surface density. These trends qualitatively agree with expectations for carbon and oxygen isotopic abundance variations due to stellar nucleosynthesis, with a possible effect of fractionation.

\end{abstract}

\renewcommand{\thefootnote}{\fnsymbol{footnote}}


\section{Introduction}
\label{intro}

Rotational transitions of the isotopologues of $^{12}$CO are observable as discrete lines, with the strongest being $^{13}$CO and C$^{18}$O. Although these lines are fainter than $^{12}$CO, they can be useful diagnostic tools. In contrast to the low-$J$ $^{12}$CO lines, $^{13}$CO and C$^{18}$O transitions often remain optically thin over large areas in galaxies. Contrasting them with the thick $^{12}$CO lines can constrain the optical depth of both lines and illuminate the underlying molecular gas column and volume densities \citep[e.g.][]{1982ApJ...258..467Y, 2008ApJ...679..481P, 2009tra..book.....W}. Comparing transitions of different optically thin isotopologues, e.g., $^{13}$CO/C$^{18}$O, offers the chance to trace  abundance variations across the disks of galaxies. Such variations may be due to chemistry and/or stellar nucleosynthesis. 

The main obstacle to studying the rare CO isotopologues is the faintness of their emission lines, which is driven by their lower abundance. Typically $^{13}$CO and C$^{18}$O are $\sim50$ and $\sim500$ times less abundant than $^{12}$CO. As a consequence, these isotopologues have been mainly studied in bright, actively star forming systems such as (ultra)luminous infrared galaxies (U/LIRGs) and starburst galaxy centers \citep[e.g.][]{2004AJ....127.2069M, 2011A&A...528A..30C, 2013A&A...549A..39A}. Most of our knowledge about the relative variation of the $^{13}$CO and C$^{18}$O lines across the disk of a normal star-forming galaxy comes from the Milky Way \citep{1990ApJ...357..477L, 1994ARA&A..32..191W}. Better knowledge of this ratio across the disks of normal galaxies will inform our interpretation of $^{12}$CO emission, cloud chemistry, and the influence of recent nucleosynthesis.

In this {\em Letter}, we report measurements of the $^{13}$CO~(1-0)-to-C$^{18}$O~(1-0) ratio across wide areas in the disks of nine nearby galaxies. These are targets of the IRAM large program EMPIRE (EMIR Multiline Probe of the ISM Regulating Galaxy Evolution; \citet{2016ApJ...822L..26B}) and a related ALMA program (Gallagher et al., in prep., \citet{jimenez}).

\begin{figure*}
\centering
\includegraphics[scale=0.70, angle=270]{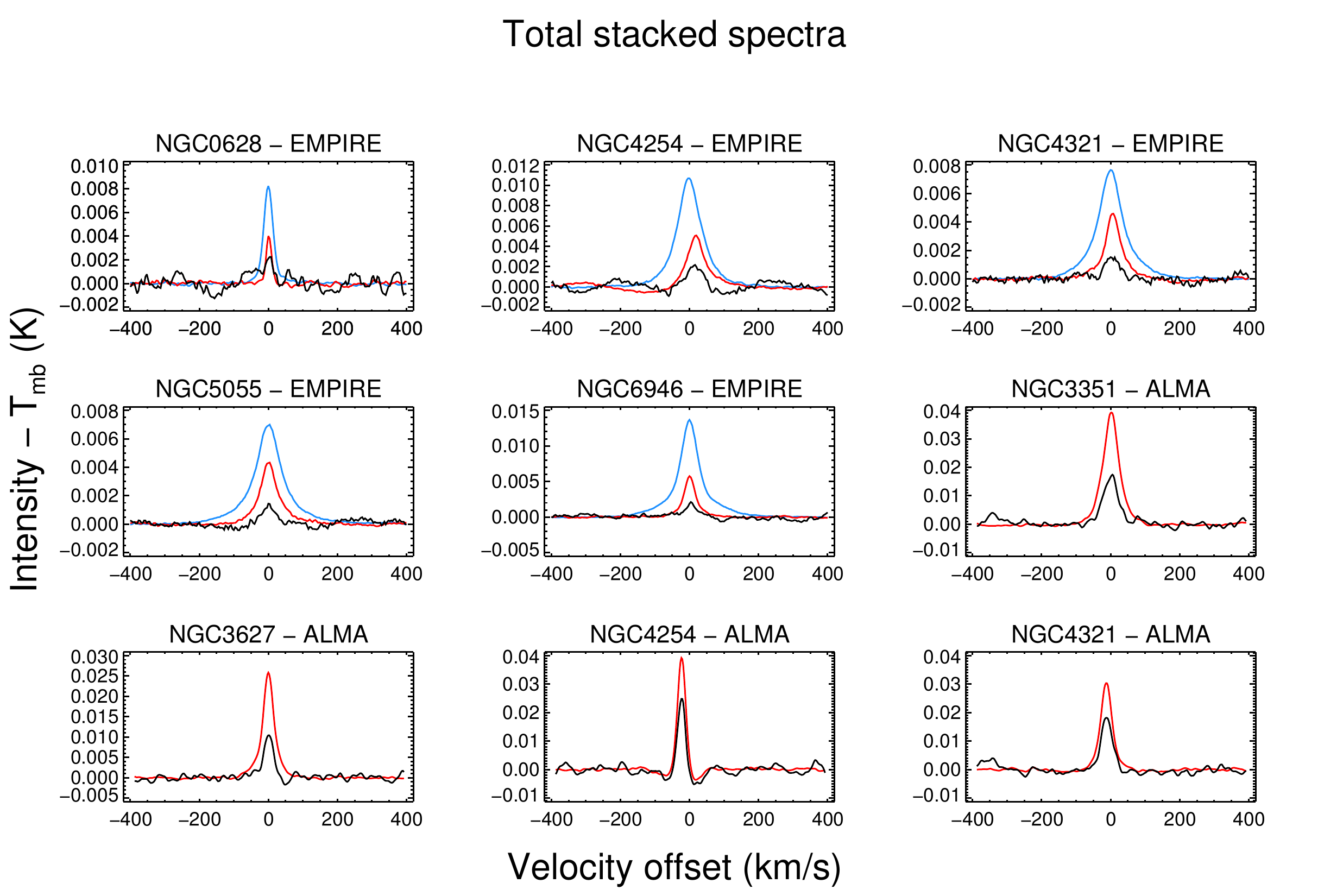}

\caption{Stacked spectra for $^{12}$CO(2-1) (blue), $^{13}$CO(1-0) (red) and C$^{18}$O(1-0) (black) emission summing over the whole IRAM 30-m and ALMA maps. C$^{18}$O is securely detected across seven targets (NGC~628, NGC~3351, NGC~3627, NGC~4254, NGC~4321, NGC~5055 and NGC~6946). The $^{13}$CO spectra are scaled down by a factor of 3 for comparison, while the $^{12}$CO spectra are scaled down by a factor of 10.}

\label{fig:total}
\end{figure*}

\begin{figure*}
\centering
\includegraphics[scale=0.4]{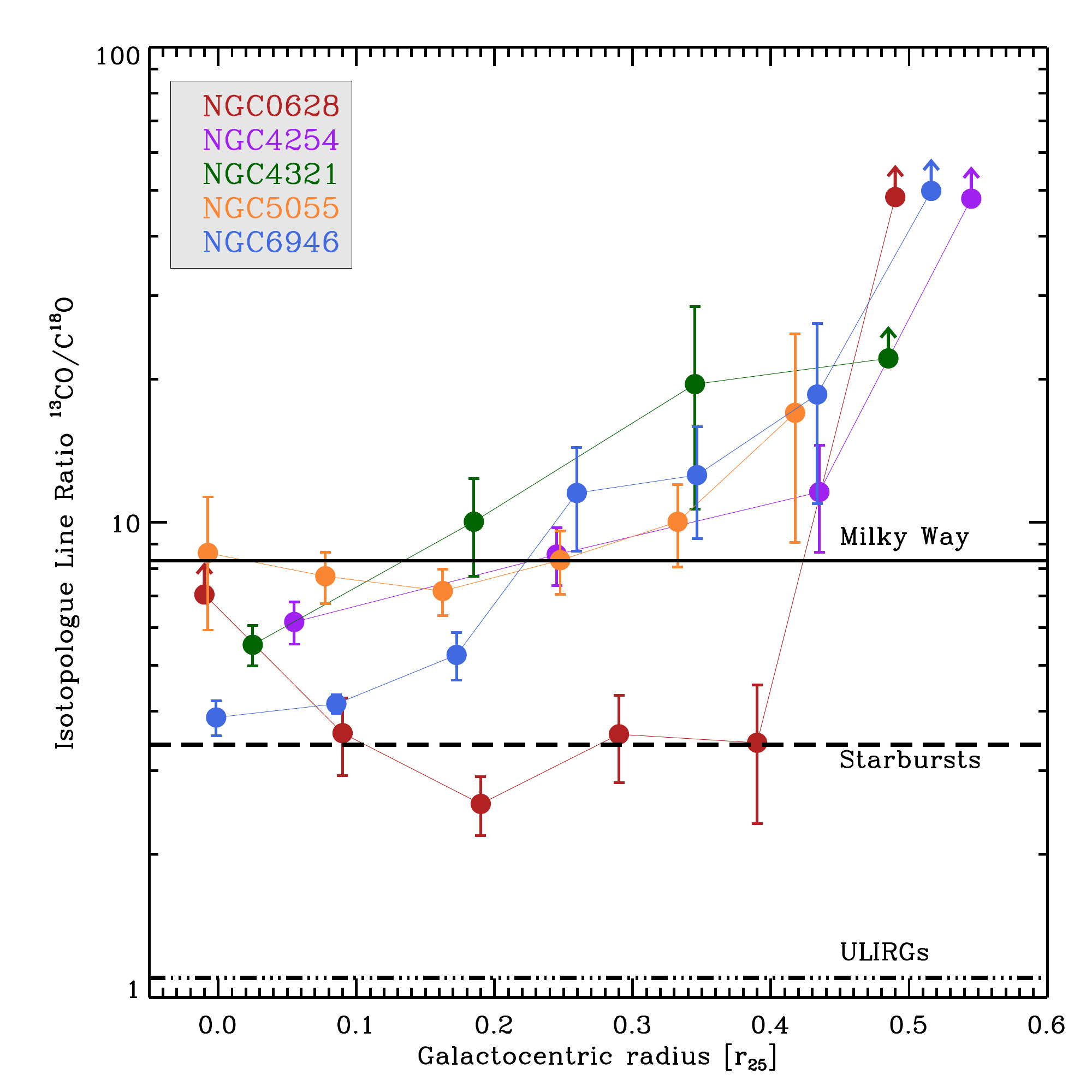}\quad
\includegraphics[scale=0.4]{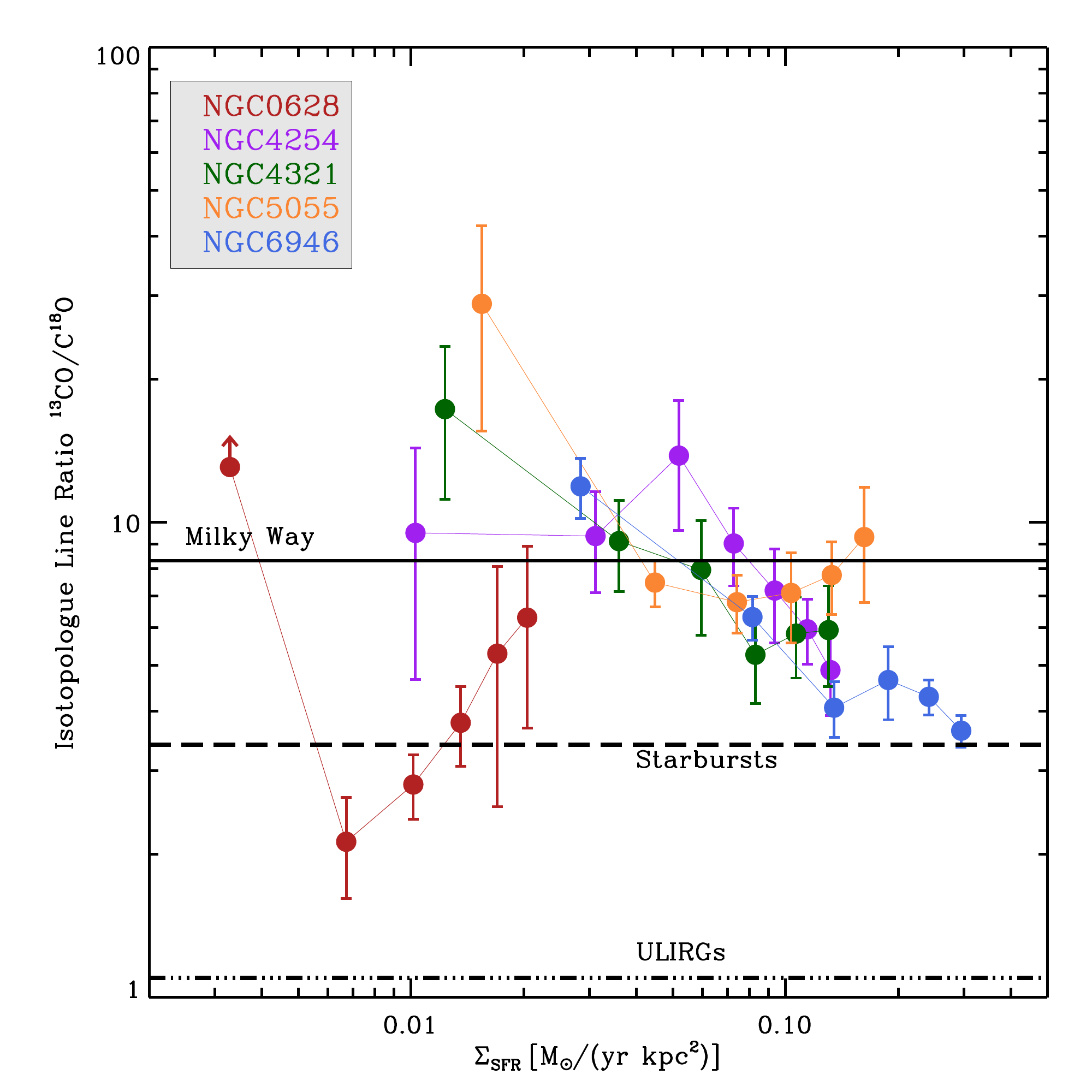}
\caption{Stacked isotopologue line ratio $^{13}$CO/C$^{18}$O as a function of galactocentric radius (left panel, in units of r$_{25}$) and star formation rate surface density (right panel). The panels show each measurement individually as a colored dot and the error bars show the uncertainty (1$\sigma$) from fitting the stacked spectrum. For those radial bins where C$^{18}$O is not securely detected, we compute upper limits to the C$^{18}$O emission (plotted as arrows). The radial profiles are derived for 30" bins for all galaxies (left panel) and in bins of 0.2 times the maximum $\Sigma_{\rm SFR}$ value for each galaxy (right panel). The black lines show the mean values for Milky Way, starburst galaxies and ULIRGS from Table \ref{table_values}.}
\vspace{0.2in}
\label{fig:trends}
\end{figure*}

\begin{figure*}
\plottwo{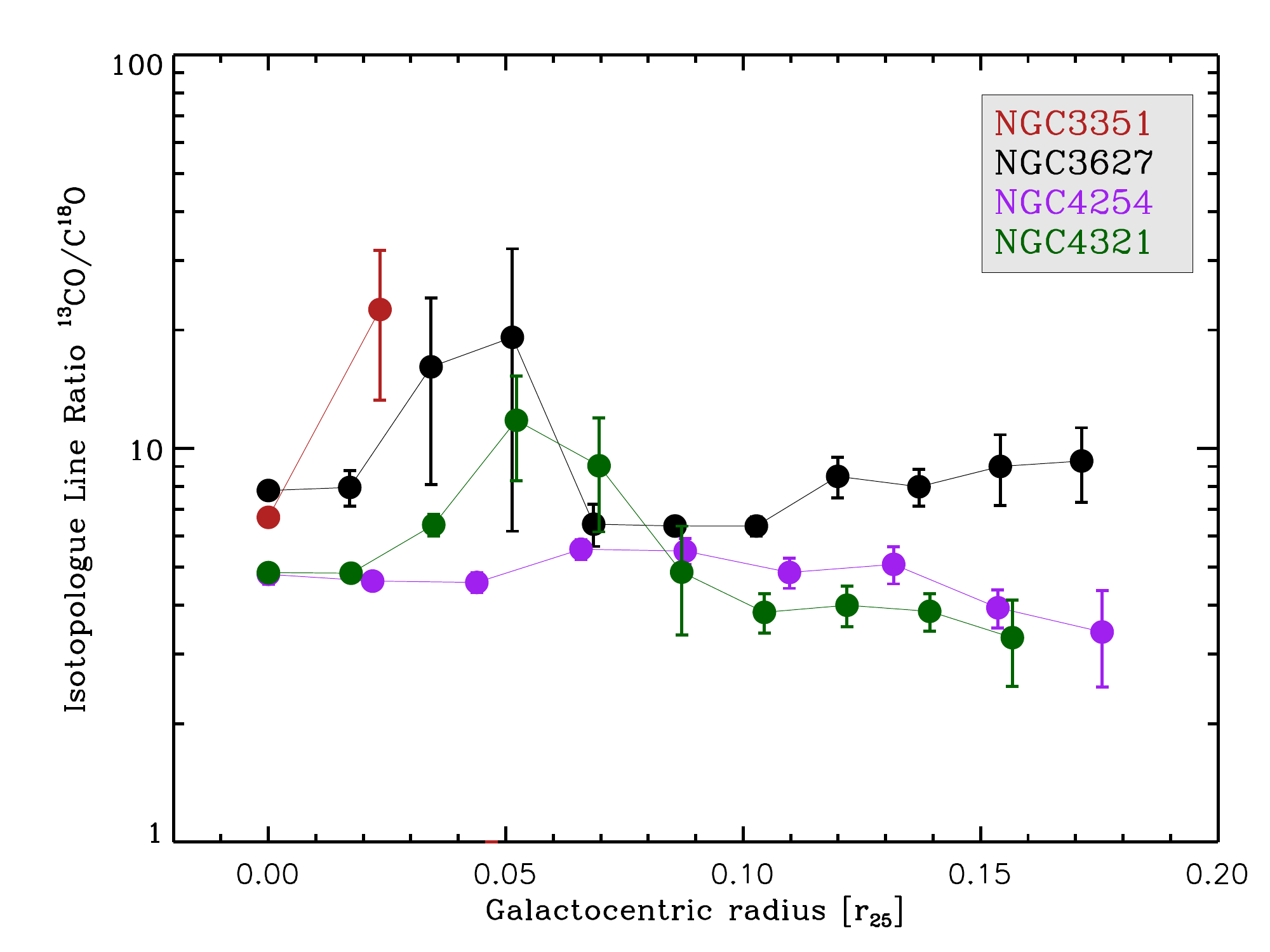}{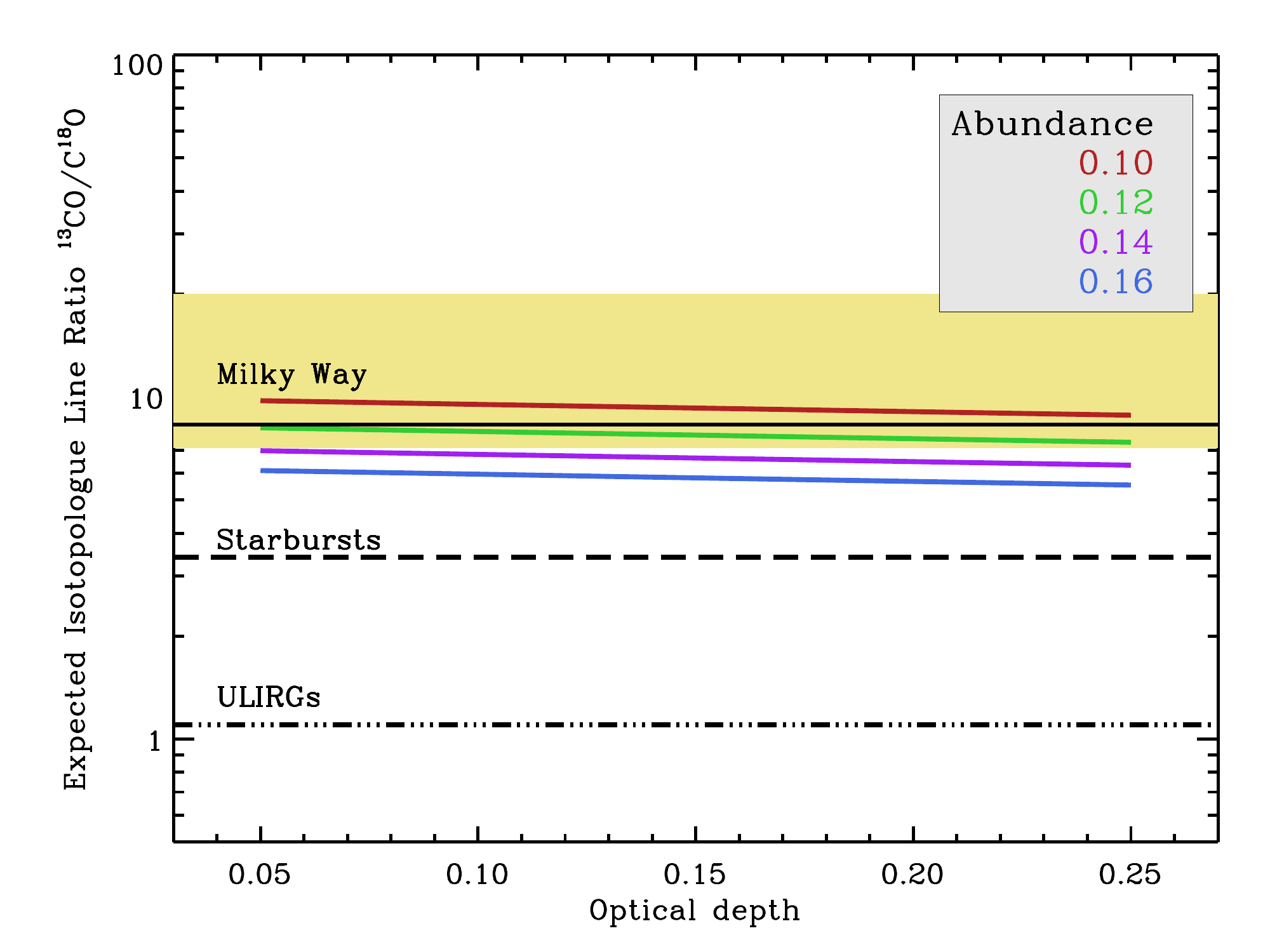}
\caption{Left: Stacked $^{13}$CO/C$^{18}$O ratios of the inner disk at high-resolution for the four galaxies that have ALMA data vs. radius. The error bars reflect the 1$\sigma$ uncertainty from the fit to the stacked spectra. Right: Expected $^{13}$CO/C$^{18}$O line ratio dependence on optical depth for 4 different abundance ratios over the range of modest $\tau_{\rm 13CO}$ that we find. The effect of changing optical depth in this regime is expected to be negligible. The yellow area represents the region covered by our observations.}
\vspace{0.2in}
\label{fig:alma_rad}
\end{figure*}

\section{Observations}
\label{sec:data}

EMPIRE \citep{2016ApJ...822L..26B} is mapping a suite of molecular lines across the whole area of nine nearby galaxy disks using the IRAM 30-m telescope. This includes $^{13}$CO~(1-0) and C$^{18}$O~(1-0), with rest frequencies of 110.20 and 109.78\,GHz. The observations began in December of 2014 and are still underway. Science-ready $^{13}$CO and C$^{18}$O data now exist for the nine galaxies listed in Table \ref{table:sources}.

\begin{deluxetable}{l c c c c c c}
\tablecaption{Adopted Properties of Observed Galaxies \label{table:sources}}      
\tablehead{
\colhead{Source} &
\colhead{Distance} &
\colhead{Incl.} &
\colhead{P.A.} &
\colhead{$r_{25}$} &
\colhead{Type} &
\colhead{$\Sigma_{\text{SFR}}$} \\
\colhead{} &
\colhead{(Mpc)} &
\colhead{($^{\circ}$)} &
\colhead{($^{\circ}$)} &
\colhead{(kpc)} &
\colhead{} &
\colhead{(\tablenotemark{d})}}
\startdata
NGC\,628\tablenotemark{a} & 9.0 & 7 & 20 & 13.1 & SAc & 4.0\\
NGC\,2903\tablenotemark{a} & 8.7 & 65 & 204 & 15.2 & SABbc & 5.7\\
NGC\,3184\tablenotemark{a} & 12.0 & 16 & 179 & 13.0 & SABcd & 2.8\\
NGC\,3351\tablenotemark{b} & 10.1 & 11 & 73 & 10.6 & SBb & 5.2\\
NGC\,3627\tablenotemark{c} & 9.8 & 62 & 173 & 14.6 & SABb & 7.7\\
NGC\,4254\tablenotemark{c} & 15.6 & 32 & 55 & 11.4 & SAc & 18\\
NGC\,4321\tablenotemark{c} & 16.4 & 30 & 153 & 14.3 & SABbc & 9.0\\
NGC\,5055\tablenotemark{a} & 8.2 & 59 & 102 & 14.0 & SABbc & 4.1\\
NGC\,6946\tablenotemark{a} & 5.6 & 33 & 243 & 9.35 & SABcd & 21

\enddata
\tablenotetext{a}{Observed with only IRAM.}
\tablenotetext{b}{Observed with only ALMA.}
\tablenotetext{c}{Observed with both IRAM and ALMA.}
\tablenotetext{d}{Average SFR surface density inside 0.75$r_{25}$, units of $\times 10^{-3}\,M_\odot$ yr$^{-1}$ kpc$^{-2}$. From \citet{2013AJ....146...19L}.}
\end{deluxetable}

\citet{2016ApJ...822L..26B} describe the basic EMPIRE observation and reduction strategy. In brief, we observed each galaxy in on-the-fly mode using the 3~mm band of the dual-polarization EMIR receiver \citep{2012A&A...538A..89C} and the Fourier transform spectrometers. We reduce the data using the GILDAS/CLASS software, subtracting a low order baseline and flagging data with rms noise that deviates strongly from that expected based on the radiometer equation. Finally the data for each spectral line were gridded into a cube, and then we removed low-order baselines from the gridded cube. The final cubes were convolved to a common resolution of $33\arcsec$ and corrected for the main beam efficiency. At $33\arcsec$ resolution with $4$~km~s$^{-1}$ wide channels, the typical rms noise is $\sim2.0$~mK ($T_{mb}$). Line calibrators observed as part of EMPIRE varied by $\sim5{-}8$\% from observation to observation. We expect the $^{13}$CO/C$^{18}$O ratio to be even more robust because both lines were observed simultaneously and any gain variations will apply to both. Statistical errors and baseline errors induced by receiver instabilities therefore dominate the uncertainties in this paper.

We also observed four targets with ALMA, using a spectral setup that covered $^{13}$CO and C$^{18}$O. Three of the four ALMA observed systems overlap with the EMPIRE sample (NGC 3627, 4254, 4321). For this letter, we use versions of these cubes convolved to 8$\arcsec$. Those that overlap EMPIRE have been corrected to account for short and zero spacing information using the \texttt{CASA} task {\tt feather}; for more details, see Gallagher et al. (in prep.). The close frequency of the two lines ($109.8$ and $110.2$~GHz), convolution to a common resolution, and simultaneous observation of $^{13}$CO and C$^{18}$O (so that they have nearly matched $u-v$ coverage) should make the internal line ratios from ALMA robust.

Both $^{13}$CO and C$^{18}$O become faint at large galactocentric radii. In order to measure their intensities at large $r_{\rm gal}$, we employ spectral stacking. To do this, we divide the galaxies into zones of fixed radius or IR surface brightness. Then, we follow a procedure similar to that of \citet{2011AJ....142...37S} and \citet{2013AJ....146..150C}: before averaging the spectra, we estimate the local mean velocity of gas from bright $^{12}$CO~(2-1) emission from HERACLES \citep{2009AJ....137.4670L}. We recenter the $^{13}$CO and C$^{18}$O spectra about this local mean $^{12}$CO velocity. Then, we average all spectra in each zone to construct a single, high signal-to-noise spectrum. After stacking, baseline issues due to receiver and weather instabilities sometimes emerge. When needed, we correct for these with an additional low-order polynomial fit.

We use this spectrum to measure the integrated intensity of the line, summing over the channels that show bright $^{12}$CO(2-1) emission. We apply a similar procedure to the ALMA data, but using the brighter $^{13}$CO line to stack the C$^{18}$O. Thus the ALMA measurements are restricted to the regions of bright $^{13}$CO emission. We calculated the uncertainty in the integrated intensity based on the noise estimated from the signal-free part of the stacked spectrum and the width of the integration window (usually $\sim$60\,km\,s$^{-1}$).

We use the total infrared (TIR) surface brightness as a proxy for the local surface density of star formation. To estimate this, we combine $\lambda = 70$, 160, and 250$\mu$m maps from {\em Herschel} \citep[KINGFISH][]{2011PASP..123.1347K}. We convolve these to match the $33\arcsec$ beam of our line data \citep{2011PASP..123.1218A}, calculate the TIR surface brightness following \citet[][]{2013MNRAS.431.1956G}, and then convert to star formation rate surface density using the prescription of \citet{2011ApJ...737...67M}. NGC~2903 lacks {\em Herschel} data, therefore we use {\em Spitzer} 24$\mu$m \citep[from LVL,][]{2009ApJ...703..517D} to estimate the IR surface brightness.

\section{Results}
\label{sec:results}

Figure \ref{fig:total} shows spectra of both lines averaged over the whole EMPIRE or ALMA area for each target. We observe widespread $^{13}$CO(1-0) from each galaxy while C$^{18}$O(1-0) is fainter, requiring substantial averaging to achieve good signal to noise. After averaging, we detect C$^{18}$O at good significance for six targets in EMPIRE. The other two, NGC\,2903 and NGC\,3184, show only weak emission and we place lower limits on $^{13}$CO/C$^{18}$O in these systems. We recover the C$^{18}$O line in all four ALMA targets.

Table \ref{table:results} reports our measured $^{13}$CO/C$^{18}$O ratios of line-integrated intensities on the $T_\text{mb}$ scale. For each target, we quote the average ratio over the whole galaxy (from EMPIRE) and the mean ratio within the inner $30\arcsec$ radius from ALMA and EMPIRE combined (except for NGC~3351, see Section \ref{sec:data}). In cases without clear detections, we report $3\sigma$ upper limits. Treating all galaxies equally, but neglecting upper limits, we find a mean ratio of 7.9$\pm$0.8 for the whole galaxy disks. Except in NGC~628, this galaxy-averaged ratio does not vary much from galaxy to galaxy. This agrees with results by \citet{2014MNRAS.445.2378D}, who compiled observations of a large set of Seyfert, starburst, and normal star-forming nuclei and found no clear trend relating the $^{13}$CO/C$^{18}$O ratio to galaxy type.

The average $^{13}$CO/C$^{18}$O ratio for the inner $30\arcsec$ region of our targets is 6.0$\pm$0.9, slightly lower than the disk-averaged value. Our mean values for both the disk and the nuclear regions of our targets are consistent with early work on the Milky Way, which found a $^{13}$CO/C$^{18}$O abundance ratio of 5-10 \citep{1990ApJ...357..477L}, and more recent results from \citet{2008A&A...487..237W}. 

Our mean values for local star forming galaxies differ from those found for starburst and ULIRGs. Table \ref{table:results} includes a compilation of literature measurements targeting the central regions of starburst galaxies and ULIRGs. There the $^{13}$CO/C$^{18}$O ratio tends to be lower than what we find, with the lines nearly equal in strength. Our mean value for disk galaxies differs from that found for starburst galaxies by a factor of $\sim$2 \citep[e.g.][]{2011RAA....11..787T, 2014A&A...565A...3H} and from that found for ULIRGs by a factor of $\sim$6 \citep[e.g.][]{2009ApJ...692.1432G, 2013MNRAS.436.2793D}.

Averaging over whole galaxies may obscure variations in the line ratio by blending together many different physical conditions. In Figure \ref{fig:trends}, we break apart the galaxies where we securely detect C$^{18}$O at multiple radii (this removes NGC~3627 from the sample, but we show its profile using ALMA data below). We plot $^{13}$CO/C$^{18}$O as a function of galactocentric radius and star formation rate surface density.

In the left panel of Figure \ref{fig:trends}, we see $^{13}$CO/C$^{18}$O increase with increasing radius for NGC\,4254, NGC\,4321, NGC\,5055 and NGC\,6946. Again NGC~628 represents the outlier, with a low ratio and perhaps a dropping rather than rising profile. \citet{2008A&A...487..237W} showed a similar trend for $^{13}$CO/C$^{18}$O to increase with increasing radius in the Milky Way (see their Figure 3a), consistent with the EMPIRE observations of normal-disk galaxies. These rising profiles imply high $^{13}$CO/C$^{18}$O at the largest radius where we can still achieve a detection, typically $^{13}$CO/C$^{18}$O $\sim$16 at $\sim$ 0.4$\times r_{25}$. At this radius, the ISM transitions from H$_2$ to H\,I dominate \citep[see][]{2011AJ....142...37S}. To our knowledge, such high ratios have not been reported before for normal star-forming disk galaxies.

Using ALMA, we construct more detailed profiles of the inner part of four of our targets. We show these in Figure \ref{fig:alma_rad}. The inner, molecule-rich parts of these galaxies do not yet show the rising profiles seen at larger radii in Figure \ref{fig:trends}. However, these ``zoomed in'' profiles show several features not immediately visible at lower resolution. Three of our targets, NGC~3351, NGC~3627, and NGC~4321 host bright inner regions ($\sim$500\,pc, $\sim0.05\,r_{25}$) likely fed by gas flows along the bars in these galaxies. No such features are apparent in NGC 4254, which lacks a strong bar. At least in the barred galaxies we might expect less isotopic abundance variations due to efficient mixing of the gas in their inner parts, similar to what is observed in the Milky Way.

The right panel in Figure \ref{fig:trends} shows the $^{13}$CO-to-C$^{18}$O ratio as a function of the surface density of star formation, $\Sigma_{\rm SFR}$. Again, NGC 628 appears as an outlier, while the other targets with resolved gradients show decreasing $^{13}$CO/C$^{18}$O with increasing $\Sigma_{\rm SFR}$. As discussed below, the sense of this trend is what is expected if massive star nucleosynthesis alters the isotopic abundances, but it might also reflect optical depth effects. In either case, as we sort our targets by $\Sigma_{\rm SFR}$, we observe a clear trend from ``Milky Way'' type values in low $\Sigma_{\rm SFR}$ regions towards ``Starburst'' type values in high $\Sigma_{\rm SFR}$ regions, though our targets do not approach the ULIRG regime. We discuss explanations for these trend in the next section.

\begin{deluxetable}{l c c}
\tablecaption{Measured $^{13}$CO/C$^{18}$O line intensity ratios for our sample and literature values \label{table:results}}      
\tablehead{
\colhead{Source} &
\colhead{Total disk} &
\colhead{Central 30"}\\
\colhead{} & \colhead{(EMPIRE)} & \colhead{(ALMA \& EMPIRE)}}
\startdata
{\bf EMPIRE \&ALMA} & & \\
\hline
NGC 628 & 2.4$\pm$0.8 & $>$ 2.5\\
NGC 2903 & $>$ 8.0 & $>$ 7.0\\
NGC 3184 & $>$ 8.2 & $>$ 6.2\\
NGC 3351\tablenotemark{*} & -- & 5.4$\pm$0.8\\
NGC 3627 & 9.0$\pm$1.1 & $>$ 3.2\\
NGC 4254 & 8.5$\pm$0.6 & 6.2$\pm$0.7\\
NGC 4321 & 9.8$\pm$0.5 & 5.4$\pm$0.7\\
NGC 5055 & 9.9$\pm$0.8 & 8.7$\pm$2.5\\
NGC 6946 & 7.6$\pm$0.5 & 3.8$\pm$0.3\\

\hline                    
{\bf LITERATURE} & & \\
\hline   
Milky Way\tablenotemark{a} & 8.27$\pm$0.2 & 7.1$\pm$0.2\\
Solar System\tablenotemark{b} & 5.5 & -- \\
LMC\tablenotemark{c} & 30$\pm$5 & --\\
NGC\,5194 & 2.6$\pm$1.7\tablenotemark{d} & 3.6$\pm$0.3\tablenotemark{e}\\
NGC 6946\tablenotemark{f} & -- & 2.3$\pm$0.2\\
M82\tablenotemark{d} -- Starburst & -- & 2.7$\pm$0.9\\
NGC\,253\tablenotemark{g} -- Starburst & -- & 3.60$\pm$0.04\\
Maffei2\tablenotemark{h} -- Starburst & -- & 3.8$\pm$0.8\\
NGC\,1068\tablenotemark{i} -- Starburst & -- & 3.4$\pm$0.9\\
Arp 220\tablenotemark{j} -- ULIRG & -- & 1.0$\pm$0.4\\
Mrk 231\tablenotemark{g} -- ULIRG & -- & 1.3$\pm$0.4\\
SMM J2135\tablenotemark{k} -- High-$z$ & -- & 1.0$\pm$0.3
\enddata
\tablenotetext{*}{Only observed with ALMA, measurements for NGC~3351 cover the central $\sim$1~kpc ($\sim$20").}
\tablenotetext{a}{\citet{2008A&A...487..237W}}
\tablenotetext{b}{\citet{1994ARA&A..32..191W}}
\tablenotetext{c}{\citet{1998A&A...332..493H}}
\tablenotetext{d}{\citet{2011RAA....11..787T}}
\tablenotetext{e}{\citet{2008PASJ...60.1231V}}
\tablenotetext{f}{\citet{2004AJ....127.2069M}}
\tablenotetext{g}{\citet{2014A&A...565A...3H}}
\tablenotetext{h}{\citet{2008ApJ...675..281M}}
\tablenotetext{i}{\citet{2013A&A...549A..39A}}
\tablenotetext{j}{\citet{2009ApJ...692.1432G}}
\tablenotetext{k}{\citet{2013MNRAS.436.2793D}}
\label{table_values}
\end{deluxetable}

\section{Discussion}
\label{sec:summary}

Leaving aside NGC 628, our targets tend to show rising $^{13}$CO/C$^{18}$O with increasing galactocentric radius and falling $^{13}$CO/C$^{18}$O with increasing $\Sigma_{\rm SFR}$. What drives these variations? We consider two scenarios that could explain our observations: (1) changes in the abundances of the observed molecules, and (2) changes in the optical depths of the observed lines.

\subsection{Variations in the Molecular Abundances}

The observed ratio $^{13}$CO/C$^{18}$O will vary if the relative abundance of the two species varies. We highlight selective photodissociation, isotope dependent fractionation, and selective enrichment by massive star nucleosynthesis as possible explanations.

Ultraviolet radiation dissociates the molecules in molecular clouds, and preferential dissociation of one species or another could change the $^{13}$CO/C$^{18}$O ratio. Selective photodissociation is possible because $^{13}$CO is more effective at self-shielding against UV photons than C$^{18}$O due to its higher abundance \citep{1982ApJ...255..143B}. However, even for the far more abundant isotopolgue $^{12}$CO, \citet{2017MNRAS.465..885S} find that self-shielding is only comparable in importance to shielding by dust and cross-shielding by the Lyman-Werner damping wings of H$_2$. For the much less abundant $^{13}$CO and C$^{18}$O, self-shielding must therefore be unimportant compared to dust- and H$_2$-shielding, which are non-selective. Indeed, using \texttt{DESPOTIC} \citep{2014MNRAS.437.1662K}, we find that these mechanisms together are sufficient to reduce the dissociation rate for $^{12}$CO and its isotopologues to near zero for columns of $N_{\text{H}_2} \sim 10^{22}$ cm$^{-2}$, even if there is no self-shielding. We can therefore discard the possibility of selective photodissociation as an explanation for our results.

\medskip

Alternatively, isotope dependent fractionation may lead to different abundances. In cold regions, where ion-molecule chemistry dominates, there would be preferential formation of $^{13}$CO by chemical fractionation:

\begin{equation}
\label{fractionation}
^{13}\textnormal{C}^+ + ^{12}\textnormal{CO} \rightarrow ^{12}\textnormal{C}^+ + ^{13}\textnormal{CO} + \Delta \textnormal{E}
\end{equation}   

If, as expected, cloud temperature scales with $\Sigma_{\rm SFR}$, and the C$^{18}$O abundance is not affected by fractionation processes, then the $^{13}$CO/C$^{18}$O ratio will decrease with $\Sigma_{\rm SFR}$ and will increase with the distance to the center of the galaxy. This could explain the increasing radial profiles for the sample if the gas is cold enough further out in their disks, as well as the increasing profiles seen in Figure \ref{fig:trends} as a function of $\Sigma_{\rm SFR}$. It would not explain the decrement in the $^{13}$CO/C$^{18}$O ratio of NGC\,628 as a function of $\Sigma_{\rm SFR}$.

Very massive stars produce $^{12}$C at the end of their lives. The $^{18}$O is also produced by high-mass stars, but the yield depends on the amount of $^{12}$C and $^{16}$O that is available \citep[e.g.][and references within]{1993A&A...274..730H, 2004AJ....127.2069M, 2013ARA&A..51..457N}. On the other hand, $^{13}$C is only produced as a result from the CN cycle of Helium Burning in intermediate-mass stars, or as a secondary product in low and high mass ($>$10 M$_\odot$) stars \citep{1991A&A...249...31S}. Thus, one would expect recent star formation to increase the $^{18}$O and $^{12}$C abundances on a short timescale, as massive stars explode as supernovae and enrich the ISM. Meanwhile, $^{13}$C comes primarily from the red giant phase of intermediate-mass stars and requires more time to enrich the ISM. C$^{18}$O could thus be expected to become overabundant relative to $^{13}$CO in regions where enrichment from young stars sets the isotopic abundances. In reality, the abundances in the molecular gas will reflect a complex combination of pre-existing enrichment, enrichment from recent star formation, and removal of gas by feedback. But the basic explanation is that gas preferentially enriched by massive stars should show low $^{13}$CO/C$^{18}$O.

Abundance variations due to selective nucleosynthesis offer a plausible mechanism that explains the increasing radial profiles of our sample. However, the observed $^{12}$C/$^{13}$C abundance ratio in the Milky Way shows the opposite trend \citep{1990ApJ...357..477L, 2005ApJ...634.1126M}, implying that the $^{16}$O/$^{18}$O abundance ratio decreases with galactocentric radius as well. This argues against nucleosynthesis as a general explanation. In NGC\,628 where the radial profiles are flat, there could be older populations where the $^{13}$C has been created more efficiently and well mixed back into the ISM afterwards. This galaxy could also present strong fractionation effects; however more lines are needed to investigate it in detail.

\subsection{Changes in the Optical Depth}

Our observations could also be explained if the optical depth of $^{13}$CO changes across a galaxy. We expect that $^{12}$CO remains optically thick over most areas in our targets. Variations in $^{13}$CO optical depth could result from changing line widths or changes in excitation due to temperature or density variations \citep[e.g., see][]{2004AJ....127.2069M}.

We gauge the importance of the optical depth effects on our $^{13}$CO measurements by comparing to $^{12}$CO measurements (Jim\'enez-Donaire et al. in prep., Cormier et al., in prep.). Assuming that the $^{12}$CO is optically thick and that both species have a common excitation temperature and beam filling factor (our observations show that $^{13}$CO and $^{12}$CO have similar large structure, similarly to \citet{2009ApJ...699.1092H}) the ratio allows us to estimate the optical depth of the $^{13}$CO transition, $\tau_{13CO}$. Applying these simple assumptions to our observations, and assuming that the CO~(2-1) measured by HERACLES is approximately thermalized with the CO~(1-0), we find $0.05<\tau_{\rm13CO}<0.25$. These values imply optically thin $^{13}$CO across our sample at the factor of $\sim 2$ level, even if our simplifying assumptions break down.

In other words, based on the observed $^{13}$CO/$^{12}$CO line ratios, which are substantially lower than $\sim 1$, the $^{13}$CO emission that we observe appears mostly optically thin. In order to explain our observed line ratio variations, the $^{13}$CO line would need to become optically thick over some regions. Thus, we do not expect that the high $^{13}$CO/C$^{18}$O ratios that we observe originate from changes in $\tau_{13CO}$. To illustrate this, the right panel in Figure \ref{fig:alma_rad} shows the expected dependence of the isotopologue ratio on the optical depth of $^{13}$CO over the range of modest $\tau_{\rm 13CO}$ that we find. The effect of changing optical depth in this regime is negligible. We instead prefer one of the abundance-based explanations described above.

\section{Conclusions}

We report observations of the 1-0 transitions of $^{13}$CO and C$^{18}$O across the disks of nine nearby galaxies from the IRAM 30m program EMPIRE and ALMA. Using a spectral stacking approach, we measured the radial variation of this ratio and its dependence on the local surface density of star formation. Because of the faintness of the C$^{18}$O line, this is the first significant sample of $^{13}$CO/C$^{18}$O radial profiles for nearby galaxy disks.

Averaging over whole galaxy disks, we find a mean value of $^{13}$CO/C$^{18}$O ratio of $7.9\pm0.8$ with no clear variations from galaxy to galaxy. In NGC~4254, NGC~4321, NGC~5055 and NGC~6946 we do observe systematic internal variations. Here $^{13}$CO/C$^{18}$O increases by $\sim$40\% with increasing radius. We also find a decreasing $^{13}$CO/C$^{18}$O ratio with increasing surface density of star formation. As a result, the central regions of our targets have a mean $^{13}$CO/C$^{18}$O of $6.0\pm0.9$, somewhat lower than the whole disk averages. This central value resembles that found in the Milky Way \citep{2008A&A...487..237W}, but differs from values for starburst galaxies (3.4$\pm$0.9) and ULIRGs (1.1$\pm$0.4).

We argue that $^{13}$CO optical depth is unlikely to drive our observed trends. Instead, we suggest that the variations in the observed ratios reflect real changes in the abundance of the two species, driven by either selective enrichment of the ISM in active regions by massive star nucleosynthesis, or a decrease in the amount of $^{13}$CO due to suppression of $^{13}$CO in hotter, higher column density environments.\\

FB, MJ and DC acknowledge support from DFG grant BI 1546/1-1. AH acknowledges support from the Centre National d'Etudes Spatiales (CNES). AU acknowledges support from Spanish MINECO grants FIS2012-32096 and
ESP2015-68964. The work of AKL is partially supported by the National Science Foundation under Grants No. 1615105 and 1615109. MRK acknowledges support from ARC DP160100695. The National Radio Astronomy Observatory is a facility of the National Science Foundation operated under cooperative agreement by Associated Universities, Inc. This paper makes use of the following ALMA data: ADS/JAO.ALMA\#2011.0.00004.SV. ALMA is a partnership of ESO (representing its member states), NSF (USA) and NINS (Japan), together with NRC (Canada), NSC and ASIAA (Taiwan), and KASI (Republic of Korea), in cooperation with the Republic of Chile. The Joint ALMA Observatory is operated by ESO, AUI/NRAO and NAOJ.

\end{document}